Enhancement in spin-torque efficiency by nonuniform spin current generated within a tapered nanopillar spin valve

P. M. Braganca[*], O. Ozatay[†], A. G. F. Garcia[+], O. J. Lee, D. C. Ralph, R. A. Buhrman

Cornell University, Ithaca NY 14853-2501

(Dated 10/09/07)

ABSTRACT

We examine the effect a spatially non-uniform spin current with a component polarized partially out of the plane has on a low saturation magnetization nanomagnet free layer. Micromagnetic simulations indicate that the spin torque efficiency acting upon the reversing nanomagnet can be enhanced through this process, resulting in faster switching with smaller currents. In doing so, we determine that micromagnetic structure within the nanomagnets can be beneficial for reversal processes. We verify this enhancement experimentally in devices with a tapered nanopillar geometry that generates a spin current polarized partly out of plane. Finally, to take even better advantage of these effects, we examine micromagnetically the benefits of a tapered three-magnetic-layer structure that further reduces reversal times while maintaining the thermal stability of the free layer.

[*] Corresponding author email: pmb32@cornell.edu
[†] Present address: Hitachi Global Storage Technologies, San Jose Research Center, San Jose CA 95135
[+] Present address: Stanford University, Stanford CA 94305

## I. Introduction

The ability of a spin-polarized current pulse to rapidly reverse the orientation of a thin film nanomagnet through the transfer of spin angular momentum has been studied extensively due to possible uses in high performance random-access magnetic memory (MRAM). However, the realization of spin torque (ST) MRAM requires that the current level for reliable and fast writing be low enough to be compatible with both scaled CMOS transistors and high-performance magnetic tunnel junctions (MTJ) employed as MRAM elements. Additional requirements for non-volatility demand that the nanomagnet have a strong enough combination of anisotropy field $H_K$ and magnetic moment $m$ so that there is a sufficient energy barrier $U_A$ opposing random thermal reversal of the nanomagnet orientation. This poses a significant challenge, since the current for ST reversal scales with $m$, making the current (density) levels for fast (< 3 ns) ST writing quite high, > 1 mA (>$10^7$ A/cm$^2$), in experiments to date[1,2,3].

Several methods have been examined to reduce the ST reversal current $I_s$. One obvious approach is to increase the spin polarization $P$ of the incident current, but this effect begins to saturate[4,5] once $P$ > 66%. At that point, the angular momentum transferred per electron with polarization transverse to $m$ becomes very close to the ideal limit, neglecting spin accumulation effects that can occur in spin valve structures[6,7,8]. Combining this approach with two reference layers bracketing the free layer can further reduce $I_s$ by up to a factor of two[9,10], but this still may not be sufficient to realize high speed nonvolatile ST-MRAM. Other strategies involve more complicated structures, such as injecting a highly localized spin-polarized current by use of a nanoconstriction[11], or using ferromagnetic multilayers where the reference and free layers are polarized out of plane[12] due to intrinsic perpendicular anisotropy. Although these approaches can reduce $I_s$, they require advances in magnetic materials or complicated fabrication processes such that their practicality has yet to be fully demonstrated.

Here we discuss simulations and experimental results demonstrating an alternative means of substantially enhancing the efficiency of spin-polarized currents driving the fast reversal of thin film nanomagnets, in a way that does not require materials development or multiple nanolithography steps. This approach utilizes ferromagnetic material with a comparatively low saturation magnetization density $M_S$ and high spin filtering properties, such as $Ni_{81}Fe_{19}$ alloy (Py), together with a device geometry utilizing a comparatively thick reference layer with tapered sidewalls. As a consequence of the geometry, the spin current generated by the reference layer is not uniformly polarized in the plane of the film, but instead has a component with substantial out-of-plane polarization (OPP) maximized near the ends of the major axis of the device. Micromagnetic simulations (MMS), as discussed below, predict a substantially reduced threshold current required for magnetic reversal, and a significant enhancement in the rate at which the reversal time decreases with current above this threshold. These simulations are supported by experimental ST pulse-switching results obtained from spin-valve nanopillar device structures designed and fabricated to enhance the OPP component of the current flowing between the reference and free layers. Our study indicates that tuning the geometry of a ST device to obtain a spatially non-uniform OPP current component is an enabling technique for the realization of ST-MRAM with reliable nanosecond writing at low current-pulse amplitudes.

## II. Spin-torque reversal

The basics of nanomagnet reversal by spin transfer in metallic multilayers are well established[4,13]. When a spin current generated by electrons passing through or reflecting from a ferromagnetic reference layer impinges on a nanomagnet, the component of the spin current transverse to the local moment of the nanomagnet is transferred to it with an efficiency that depends on the nanomagnet's spin filtering properties. If both the polarization of the incident spin current and the easy axis of the nanomagnet are in the plane of the film, the predominant average effect of the spin transfer is, depending upon the direction of current flow, to exert either an extra damping or "anti-damping" torque on the nanomagnet. In the latter case, when $I = I_c$ the spin torque initiates magnetic oscillations of the free layer. When the switching current, $I_s$, is reached, the oscillations have grown in amplitude sufficiently that the nanomagnet moment develops a net component opposite to its original easy-axis orientation, at which point the spin torque causes the nanomagnet to settle rapidly into a quiescent magnetically-reversed state.

$I_c$ can be estimated analytically by modeling the nanomagnet as uniformly polarized and by employing the standard Landau-Lifschitz-Gilbert-Slonczewski (LLGS) equation to describe the behavior of this "macrospin." When both the reference and free layers have their equilibrium moments fully in plane we have[4,14,15] $I_c^{+/-} = \left(\frac{2e}{\hbar}\right)\frac{\alpha}{\eta^{+/-}} M_s V [2\pi M_{eff} + H_{eff}]$. Here $I_c^{+/-}$, is the critical current for the onset of dynamics when the reference and free layers are nearly parallel/anti-parallel, $\alpha$ is the Gilbert damping parameter, $e$ is the electron charge, $M_s$ is the saturation magnetization of the free layer, $V$ the free layer volume, $H_{eff}$ the effective field acting on the free layer, $4\pi M_{eff}$ is its effective demagnetization field (typically $4\pi M_{eff} >> H_{eff}$), and $\eta^{(+/-)}$ is the spin torque efficiency parameter, which is $\leq 0.5$ in the absence of spin accumulation effects, and varies with the alignment angle $\theta$ between the free and reference magnets. To the extent that the macrospin model approximates the true critical current for ST reversal of a nanomagnet, the pathway for reducing switching currents is clear; maximize $\eta$, and minimize $\alpha$, $M_s$, and $V$. However, the constraint of thermal stability, which is typically taken as requiring $U_A = M_s H_K V/2 \geq 40 k_B T$, where $T$ is the device operating temperature, and materials constraints determining damping ($\alpha \geq 0.01$ for conventional MRAM materials), provide limited flexibility for optimization. One strategy, since $H_K$ scales with both $M_S$ of the nanomagnet and its thickness, is to use a thicker free layer composed of a lower $M_S$ material to maintain $U_A$, thereby lowering $I_c$ through a reduction in the demagnetization field $4\pi M_{eff}$ (assuming high spin torque efficiency is maintained).

A different approach for ST switching is to use a spin current polarized entirely perpendicular to the plane of the in-plane magnetized free layer[16,17]. In this case, the predominate effect of the spin torque is to directly force the free layer magnetization out of plane. When this effect becomes large enough relative to $H_K$, the nanomagnet begins to precess freely about the large out-of-plane demagnetization field. Macrospin modeling[15] predicts this onset to be at $I_c^\perp = \left(\frac{2e}{\hbar}\right)\frac{M_s V}{\eta(\theta=\pi/2)}\left(\frac{H_K}{2}\right)$. Reversing the spin torque after a $90^0$ rotation of the free layer and then terminating it at the $180^0$ point could result in very rapid reversal (~ 100 ps), but this requires both precise timing of the current pulse and higher amplitudes than spin currents polarized in-plane, since typically $H_K > \alpha(2\pi M_{eff})$.

In this article, we demonstrate that a significant benefit in ns reversal can be achieved with a combination of in-plane and out-of-plane polarized spin currents. By employing the macrospin approximation, it is straightforward to obtain a qualitative understanding of this effect using simulations, although to our knowledge such a combination has not been previously discussed. This involves solving the LLGS equation for a single magnetic layer with a uniform moment, where the spin-torque term used was of the form in ref. 6, with a value of $\Lambda = 1$ for the torque asymmetry parameter to directly compare to the micromagnetic simulations discussed below. Typical material parameters were used for Py: the damping constant $\alpha = 0.014$, the $T=0$ saturation magnetization of the free layer $M_s = 650$ emu/cm$^3$ (determined by superconducting quantum interference device magnetometry measurements), easy axis anisotropy field $H_k = 150$, and spin polarization[18] $P = 0.37$. These simulations show, *e.g.*, that the reversal rate of a 5 nm thick, 45x125 nm$^2$ elliptical Py nanomagnet will be enhanced by approximately 50% if the spin current (P = 37%) has its polarization 10° out of plane, in comparison to the case of an equal current that is fully in-plane polarized (IPP). This enhancement, which does not require a precisely timed pulse, occurs because the OPP component accelerates the rate at which the macrospin moment spirals out of the plane and is somewhat similar in nature to the benefit of an applied, in-plane hard-axis magnetic field applied simultaneously with an IPP current[19]. This enhancement grows with the OPP, but when the out-of-plane torque finally becomes large enough to overcome $H_k$, the effect transitions from one assisting the IPP reversal mechanism to one where the OPP current dominates, resulting in continuous precession about the demagnetization field for as long as the current is applied. We show below that when the micromagnetic behavior of nanopillar devices and of spatially non-uniform spin currents are considered, this detrimental effect can be minimized and a small OPP component can have an even greater positive effect on short pulse ST reversal than indicated by the macrospin model.

### III. Micromagnetic simulations of spin-torque reversal

While macrospin modeling provides qualitative understanding, micromagnetic simulations (MMS) give better insight into the detailed reversal behavior of nanomagnetic structures[20]. These micromagnetic simulations[21] incorporate the LLGS equation (not including a field-like torque term) at $T = 0$ with the same spin torque and material parameters as used in the macrospin simulations, with the exchange constant $A = 1.3 \times 10^{-6}$ erg cm$^{-1}$, and the volume discretized into 2.5 nm cubes for computational purposes. Static ($I = 0$) simulations of a spin valve structure are used to determine both the field required to cancel out the average dipole field exerted on the free layer by edge charges on the reference layer for the two layer structures, and to calculate the initial micromagnetic state of the free and reference layers at the dipole field. To avoid an initial state with collinear magnetic moments in the two layers, we induce an initial in-plane misalignment (~10°) by calculating the configuration with a magnetic field along the in-plane hard axis of the ellipse. This field is turned off simultaneously with the application of the current pulse for $I \neq 0$ simulations. Dynamic ($I \neq 0$) simulations include effects from magnetic interactions between the two layers and the Oersted field due to $I$. Spin torque is exerted upon both layers, with the local spin polarization of the current incident upon a layer being dependent on the local magnetization vector of the second ferromagnet, *i.e.* the current flow was assumed to be one-dimensional[6,7,8]. We treat spins classically and use the simplifying assumption that spins transmit the parallel component and reflect the antiparallel component of the local magnetization perfectly, depending on the direction the electrons traverse. This assumption requires us to use a value of $\Lambda = 1$ for the asymmetry parameter[6] to avoid false enhancement in spin torque in one reversal direction over the other. This choice, which neglects the spin accumulation effects that are expected to be present in spin valve structures, still allows for qualitative comparison of the reversal time between different device configurations.

We first consider an elliptical disk of finite thickness that has a spatially non-uniform demagnetization tensor (unlike an ellipsoid of rotation), such that the demagnetization field $4\pi M_{eff}$ decreases significantly from the center to the ends of the major axis of the disk. When properly considered by MMS, this lowers the local critical current density $J_c$ for the onset of ST excited magnetic oscillations near the ends. Zero $T$ simulations including the Slonczewski ST term[6,21] (ST-MMS) reveal that, for currents slightly above $I_s$, ST-driven oscillations grow faster at the ends of the ellipse, resulting in a reversal process that is considerably different from uniform macrospin precession. This is indicated in Fig. 1, which shows the simulated micromagnetic evolution of a single 5 nm thick Py elliptical disk with 45 nm x 125 nm$^2$ cross-sectional dimensions. In Fig. 2, we plot the reversal rate for such a nanomagnet as determined by ST-MMS for a range of currents, assuming that the polarization of the incident electrons is in-plane and uniform across the nanomagnet's surface. In comparison to a macrospin simulation employing the same material parameters and spin-transfer efficiency $\eta$, the micromagnetic calculation predicts a reduced critical current and a switching rate at larger current increased by approximately a factor of 2. ST-MMS does indicate that the enhancement is slightly lower when a typical fully-patterned, spin-valve nanopillar device structure is modeled. Then, if the reference layer is assumed to be uniformly magnetized in-plane, ST-MMS predicts switching rates as shown in Fig. 3(d), with the difference with Fig. 1 being attributed to the effect of the non-uniform component of dipole field from the reference layer in suppressing magnetization oscillations at the ends of the free layer.

This detrimental effect of the dipole field can be largely countered by choosing the reference layer geometry and material so as to generate a *spatially non-uniform* spin current with a significant OPP component. Such a spin current can be obtained by using a relatively thick (~20 nm) low-$M_s$ reference layer, so that demagnetization effects result in an out-of-plane magnetization component at the ends of the major axis of a patterned ellipse. This effect is enhanced by tapering the edges of the reference layer, which can be accomplished via directional ion beam milling during nanopillar patterning. A cross-sectional view of the $I$=0 equilibrium state of this Py-Cu-Py spin valve structure modeled with MMS is shown in Fig. 3(a). For this structure, the magnetization cants ~ 20° out of plane at the ends of the interface adjacent to the free layer, and gradually transitions to fully-in-plane near the center. Our ST-MMS calculations for magnetic reversal in this geometry include the interactions between the free and reference layers, both magnetically and by using the reference layer magnetization to determine the local current polarization acting upon the free layer, starting in the misaligned state shown in Fig. 3(b). The simulations show that the OPP component initiates large oscillations at the ends of the free layer more rapidly than with the use of a uniform IPP current for the same initial starting state. This accelerates the reversal process (see Fig. 3(c)). For reversal times in the 1-3 ns range, the ST-MMS predict a significant additional reduction in $I_s$ for the non-uniform OPP current (see Fig. 3(d)). Our simulations do indicate that the variation of the reversal rate with bias current in the micromagnetic results may not be as regular as predicted by macrospin modeling, as at certain bias currents the oscillations originating at the two ends can, due to the different directions of the OPP, momentarily oppose each other and slow down the transition to the reversed state. However, experimental results, as discussed in part below, suggest that thermal effects may reduce these interactions, and overall the effect of micromagnetic structure is to significantly enhance reversal efficiency.

### IV. Experimental demonstration of spin-torque enhancement

We confirmed these beneficial micromagnetic effects with experiments on Py-Cu-Py spin-valve nanopillar devices fabricated from thin film multilayers deposited in two different configurations. In the first, or "standard" case, the multilayer was deposited in the following sequence: 120 Cu/20 Py/12 Cu/5.5 Py/2Cu/30 Pt, where Py is $Ni_{81}Fe_{19}$ and the thicknesses are in nm. For the "inverted" case, the multilayer stack was 120 Cu/4.5 Py/12 Cu/20 Py/2Cu/30 Pt, placing the reference layer of the patterned nanopillars above the free layer rather than below it.

The nominal lateral dimensions of the elliptical nanopillar structures were 50 x 130 nm$^2$, but sidewall tapering of the device during ion milling results in inverted samples having both larger free layers and reference layers with a substantial out-of-plane magnetization component on the side adjacent to free layer (as in Fig 3(a)).

For comparison to MMS, we performed room temperature measurements to determine ST reversal probabilities as a function of current amplitude over a range of pulse widths (1-100 ns), all of which have a significant distribution due to thermal fluctuations[1,22]. Fig. 4 plots the pulse current amplitudes $I_s$ required to provide 95% reversal probability as a function of pulse width, for the two cases where the free layer of both device configurations is reversed from a state antiparallel to the reference layer to one parallel ($AP \to P$), and *vice versa* ($P \to AP$). As predicted by ST-MMS (*cf* Fig. 2), the variation of the short-pulse reversal rate with $I$ for standard devices is indeed considerably more rapid than predicted by the macrospin model when applied for the case of $P \sim 0.37$ and free layer dimensions of the standard devices. Even more notably, and also in qualitative accord with ST-MMS, the inverted devices exhibit considerably lower switching currents, and a stronger variation with current amplitude, than the standard devices, despite a free-layer volume estimated to be ~1.2 larger.

One final point to note is that we find that the asymmetry ratio of switching currents, $I_c^+/I_c^-$, is considerably less in the inverted *vs.* the normal devices, ~ 1.2 *vs.* ~ 1.6, and in both cases considerably less than predicted by one-dimensional spin transport analysis[6,7,8]. We attribute this to the reduction, by the non-uniform magnetization of the reference layer, of the spin accumulation effects that would otherwise enhance $\eta^-$ but not $\eta^+$. This negative effect on $\eta^-$ is outweighed by the overall increase in ST efficiency by OPP.

## V. Consequences and Conclusions

More extensive $T >0$ MMS analysis and experimental studies will be required to fully quantify and optimize these micromagnetic enhancement effects, but clearly they can be quite significant. To obtain the optimum enhancement, we propose to combine the benefits of the out-of-plane spin polarization with a second magnetic reference layer using the three-magnetic-layer device structure shown in Fig. 5(a). The two thick outer reference layers are anti-parallel to each other, and the sides of the top reference layer are tapered to promote out-of-plane magnetization. The free layer is a 5 nm Py elliptical disk with lateral dimensions of 45 x 125 $nm^2$ and the reference layers are Py as well. Fig. 5(b) shows the MMS predictions for switching rates for two variations of this structure. In one case, the spin polarization for each layer is taken to be 37%, as should approximately be the case for such a metallic spin valve structure. Since this structure would not exhibit a significant magnetoresistance signal, the second case assumes a tunnel junction between the free layer and the top reference layer with a tunneling spin current polarization of 66%, which results in a near ideal spin torque efficiency of $\eta = 0.46$[23]. The simulations indicate that the free layer nanomagnet is thermally stable, $U_A \sim 1.75$ eV, and in comparison to the MMS result for a simple inverted spin-valve, the three-magnetic-layer devices switch at the same rate at ~ 2x lower current. An even simpler approach would be to eliminate the taper, and employ a nanopillar structure with thick, low $M_s$ reference layers and straight sidewalls. Here, the dipole fields originating from edge charges on the two reference layers would have the same effect as tapering the sidewalls, promoting out-of-plane magnetization at the ends of both low-$M_s$ reference layers. This design does not require an ion mill process for tapering, and MMS show that the reversal is only slightly slower than the tapered device in this case, ~5-10% (not shown). Regardless of the details, our simulations and experiments suggest that an optimally designed and fabricated three-magnetic-layer hybrid tunnel-junction/spin-valve device that takes best advantage of these micromagnetic effects could be switched within a few ns with a current approaching the range required for high performance ST-MRAM. We also note that a voltage-dependent field-like contribution to the spin torque that has been found to be significant in tunnel junctions[23] should augment the OPP effect due to the micromagnetics of this proposed structure, leading to an even more efficient ST reversal process.

## Acknowledgements

The authors would like to thank J. C. Sankey for providing us with the macrospin simulation used in this work. We would also like to thank V. S. Pribiag for useful conversations and particularly for assistance in implementing the micromagnetic code used here. This work was supported in part by the Semiconductor Research Corporation (SRC), by the Office of Naval Research (ONR), by the National Science Foundation/Nanoscale Science and Engineering Center (NSF/NSEC) program through the Cornell Center for Nanoscale Systems and by an IBM-Faculty Partnership award. This work was performed in part at the Cornell NanoScale Facility, a member of the National Nanotechnology Infrastructure Network (NNIN), which is supported by the National Science Foundation (Grant ECS 03-35765) and benefited from use of the facilities of the Cornell Center for Materials Research, which is supported by the National Science Foundation/Materials Research Science and Engineering Center (NSF/MRSEC) program.

**Figure Captions:**

**FIG. 1:** (color online) Frames of a *T=0* micromagnetic simulation of a single 5 nm thick, 45x125 nm$^2$ elliptical nanomagnet with *I* = 1.5 mA, using material parameters for Py.  Here, the free layer magnetization ($\hat{m}_{free}$) is initially at 170° to the polarization of the incident current ($\hat{m}_p$), which is uniform and in plane along the long axis of the ellipse (inset Fig. 2).  Unlike the macrospin picture, which relies on a gradually building coherent oscillation of the entire magnet, we see significant oscillations begin at the edges of the magnet, since the demagnetization field of the magnet is ~ 30% smaller there.  These oscillations drag the interior along, due to exchange interactions, leading to a relatively incoherent reversal.

**FIG. 2:** (color online) Comparison of reversal times $\tau$ for the elliptical nanomagnet discussed in Fig. 1 treated both as a macrospin and micromagnetically.  The incoherent reversal mechanism shown in Fig 1 is more efficient than coherent *T=0* macrospin reversal.  The lines are least squares fit through the simulation results.

**FIG 3:** (color online) (a) *T=0* equilibrium state of a two-layer structure with a tapered reference layer above the free layer, as calculated with MMS.  (b) Misaligned (~10°) state of the adjacent reference and free layer interfaces calculated by MMS assuming the tapered device geometry and the existence of a 200 Oe in-plane hard-axis magnetic field.  This misaligned state is used as the initial configuration for the dynamic simulations, to avoid artifacts associated with a nearly collinear initial state.  For the configuration in (b), the magnetization near the edges of the reference layer curls significantly out of plane, which generates current with a partial OPP

component and enhances the oscillations at the edges of the free layer. The amplitude of these oscillations grows quickly with the assistance of this non-uniform polarization, leading to a significantly faster reversal than with a uniform in plane polarization along the easy axis, as seen in (c), which shows the evolution of the average free-layer $M_x$ with time at 1 mA. Because the reversal process starts at the ends of the major axis of the free layer and then spreads to involve the entire nanostructure through the exchange interaction, the amplitude of $M_x$ does not grow monotonically until the reversal point as it does in the macrospin model. (d) The rates for $AP \rightarrow P$ reversal predicted by ST-MMS for the spatially non-uniform OPP case are enhanced compared to the results assuming a uniform, in-plane fixed-layer magnetization along the easy axis. The lines are least-square fits through the ST-MMS results, which deviate from linear behavior due to the incoherent nature of the reversal.

**FIG 4:** (color online) We compare reversal rate vs. current for two different device structures, with the reference layer either above (inverted device) or below (standard device) the free layer. In both cases, the sidewalls were tapered during the ion milling required for nanopillar definition. The standard-structure free layer is 5.5 nm thick with a nominal 50 x 130 nm$^2$ elliptical area, while the inverted-structure free layer is 4.5 nm thick with an area ~ 1.5 that of standard structure. We measured both the (a) $P \rightarrow AP$ and (b) $AP \rightarrow P$ reversal probability for current pulses between 1-100 ns long, and for a given pulse length we define the reversal current as the value which first achieves reversal 95% of the time. A large enhancement occurs for the inverted structure, despite the larger free layer volume.

**FIG. 5:** (color online) (a) Schematic of the equilibrium state ($T=0$) of a device with two reference layers sandwiching the free layer. Tapering the top of the structure combines the advantage of spin-torque enhancement through micromagnetic effects with added torque from the second reference layer. The spacers can be either non-magnetic metals such as Cu, or if higher polarization is required, tunnel barriers such as AlOx or MgO. (b) $T=0$ ST-MMS of 2 structures, the first with both spacers being metallic ($P = 0.37$ for both reference layers), and the second with the top spacer assumed to be a tunnel barrier ($P_1 = 0.37$, $P_2 = 0.66$). Results for the simple inverted spin valve structure ($P = 0.37$) shown in Fig. 3(a) are included for comparison. Similarly to the results in Fig. 3, the simulations begin with an initial state in which the average magnetizations of the layers are misaligned (~10°), calculated by assuming a 200 Oe magnetic field along the in-plane, hard axis. The reversal currents predicted are promising for future MRAM applications.

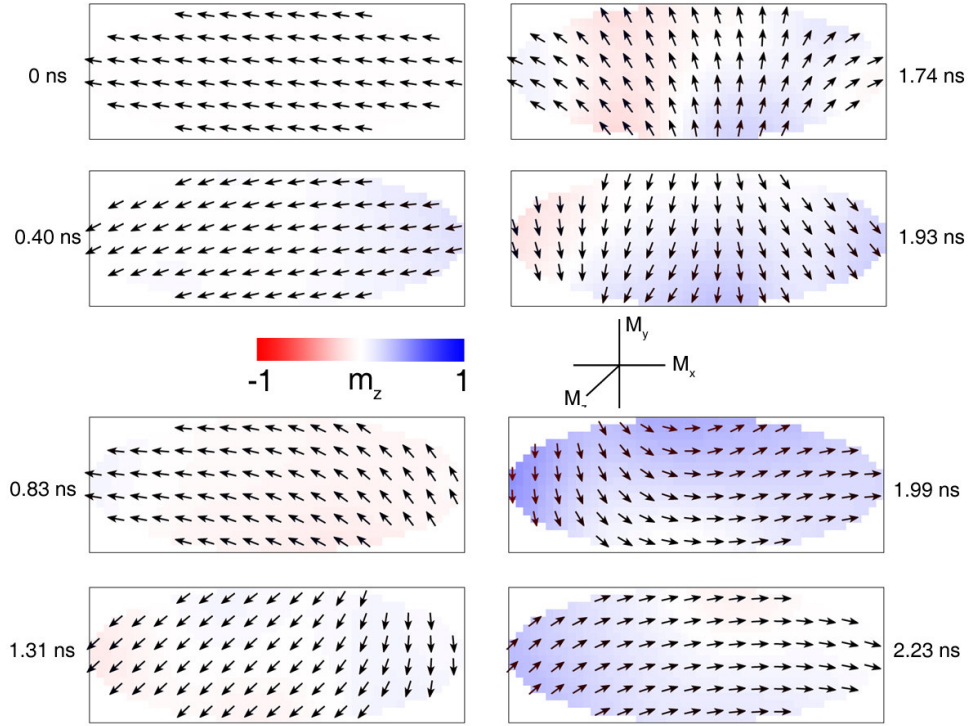

**Fig. 1**

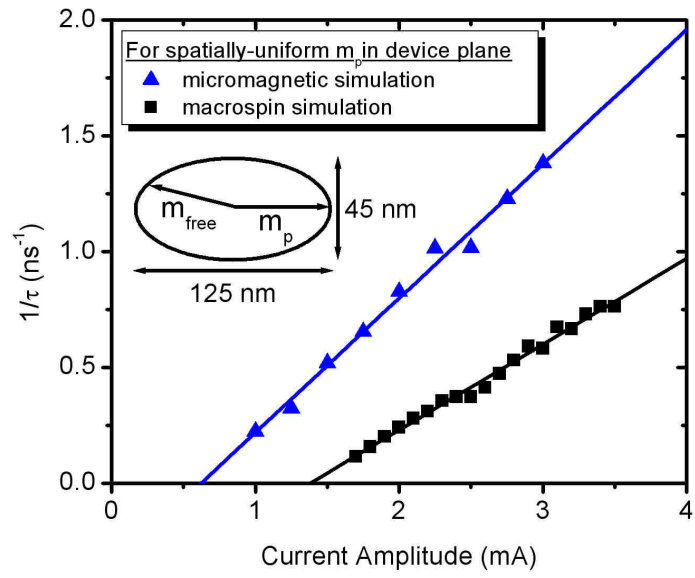

**Fig. 2**

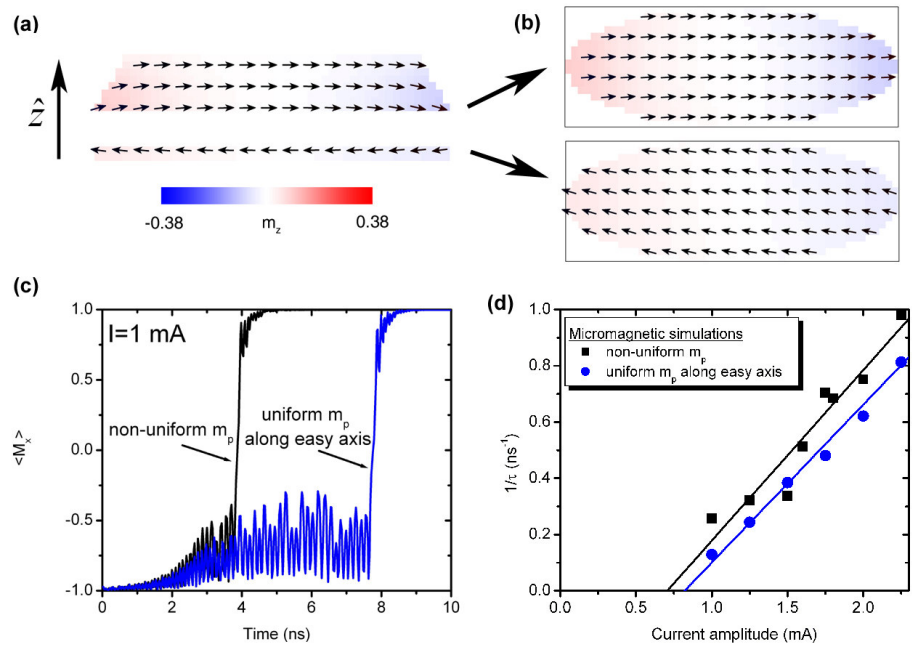

**Fig. 3**

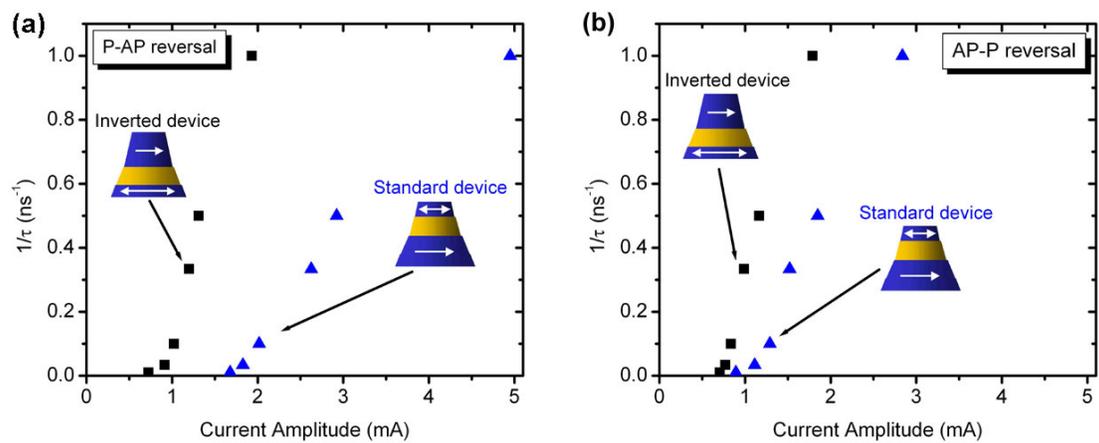

**Fig. 4**

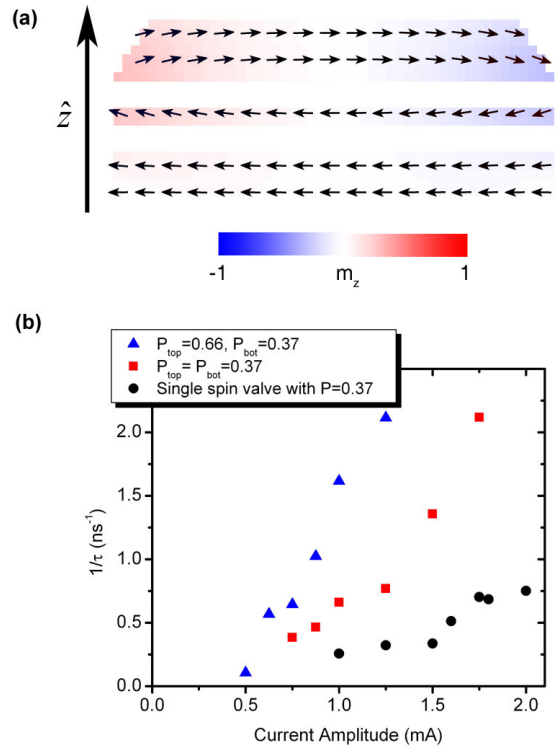

**Fig. 5**